\documentclass[fleqn,12pt,twoside]{article}
\usepackage{espcrc1}

\usepackage{epsfig}

\newcommand{\open}{< \kern -0.3em \scriptscriptstyle{)}}

\title{T-odd fragmentation functions}

\author{M. Radici\address[PV]{Dipartimento di Fisica Nucleare e Teorica, 
Universit\`a di Pavia, and \\
Istituto Nazionale di Fisica Nucleare, Sezione di Pavia, I-27100 Pavia, Italy}}
       
\begin{document}

\maketitle

\begin{abstract}
We discuss the properties of fragmentation functions arising from the interference 
of two leading hadrons produced inside the same jet in the current fragmentation
region of a hard process. For the case of semi-inclusive Deep Inelastic Scattering
(DIS), a proper folding of the cross section, integrated over the azimuthal position
of the detected hadrons, produces a factorized form that allows for the extraction 
of the quark transversity distribution at leading twist. Using an extended spectator 
model, explicit calculations are shown for the hadron pair being two pions with 
invariant mass inside the $\rho$ resonance width.
\end{abstract}

\vspace{.8cm}

Fragmentation functions (FF), like distribution functions (DF), describe the
nonperturbative properties of partons when confined inside hadronic bound states at a
low scale. Factorization theorems for hard processes, whenever available, ensure
their univsersality. Several DF have been experimentally extracted and are
parametrized with high accuracy. However, one of them, the quark transversity $h_1$,
is still unknown because it is connected to soft processes that flip chirality; as 
such, it is, e.g., unaccessible in inclusive DIS. Among alternative measurements, the
single spin asymmetries in semi-inclusive DIS seem more favourable. In fact, a new 
generation of machines (HERMES, COMPASS, eRHIC) allows for a better resolution in 
the final state and precise semi-inclusive measurements are becoming feasible. In 
this context, naive time-reversal odd (for brevity, ``T-odd'') FF naturally arise 
because no constraints from time-reversal invariance can be imposed due to the 
existence of Final State Interactions (FSI) with or inside the residual 
jet~\cite{ruju}. Some of these T-odd FF are also chiral odd, and they can be shown to
represent the natural partner to isolate $h_1$ already at leading twist. In a 
field-theoretical description that assumes factorization, the soft processes linking 
the fragmenting quark to the observed hadron(s) are defined as matrix elements of 
nonlocal operators involving quarks and gluons~\cite{collsop}. The simplest one at 
leading twist is the quark-quark correlator describing the decay of a quark $k$ into
a detected hadron $P_h$. If the transverse momentum $\vec P_{h\perp}$ is measured 
with respect to the 3-momentum transfer $\vec q \parallel \hat z$ and the quark is
transversely polarized, a T-odd chiral-odd FF arises and allows for the extraction 
of $h_1$ via a single spin asymmetry (the socalled Collins effect)~\cite{coll}. 
However, due to the lack of collinear factorization the soft-gluon radiation needs 
to be taken into account and can lead to a severe suppression of the 
effect~\cite{boer}. Moreover, knowledge of the Collins function implies 
the difficult task of modelling FSI between the observed hadron and the residual 
jet~\cite{bacc}. 

From the theoretical point of view, it is more convenient to select processes where 
two leading hadrons $P_1$ and $P_2$ ($P_1+P_2=P_h$) are detected in the same jet, 
that acts as a spectator~\cite{jaffe,noi1}. In this case, collinear factorization
holds and the cross section would not be suppressed by Sudakov form factors; 
moreover, FSI are more easily modelled inside the pair. If the two hadrons are 
unpolarized, four FF appear at leading twist: $D_1, G_1^\perp, H_1^\perp, 
H_1^{\open}$~\cite{noi1}. They depend on the light-cone quark momentum fractions 
$z_1,z_2$ delivered to the hadrons ($z=(P_1+P_2)^-/k^-=z_1+z_2$), on the transverse 
relative momentum $\vec R_\perp^2$ (with $R=(P_1-P_2)/2$), on $\vec k_\perp^2$ and 
$\vec k_\perp \cdot \vec R_\perp$~\cite{noi1}. Each FF is also related to a specific 
spin state of the fragmenting quark: $H_1^\perp$ is the analogue of the Collins 
effect; on the contrary, $H_1^{\open}$ represents a genuine new effect relating the 
transverse polarization of the fragmenting quark ($\vec S_\perp$) to the transverse 
relative dynamics of the detected pair ($\vec R_\perp$), i.e. it is an ``analyzing 
power'' that transforms $\vec S_\perp$ into the relative orbital angular momentum of 
the pair. $G_1^\perp, H_1^\perp, H_1^{\open}$ are T-odd and are nonvanishing only in 
the presence of residual FSI, at least between the two hadrons. Both $H_1^\perp, 
H_1^{\open}$ are also chiral odd and can be identified as the chiral partner needed 
to access the transversity $h_1$~\cite{noi1}. 

\vspace{-.5cm}
\begin{figure}[ht]
\begin{center}
\epsfig{file=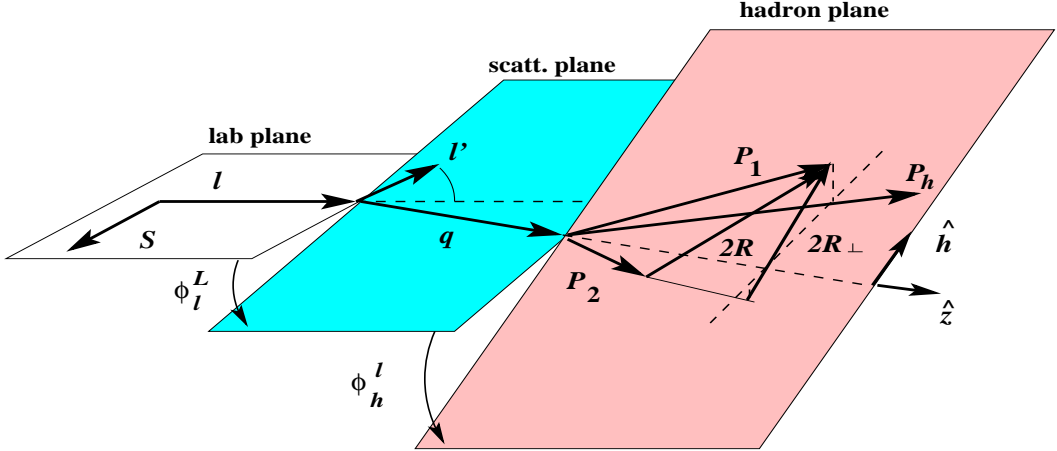,width=14cm,
height=6cm}
\end{center}
\vspace{-1cm}
\caption{Kinematics for semi-inclusive DIS where two leading hadrons are detected
(see text). \label{fig1}}
\end{figure}
\vspace{-.5cm}

The cross section at leading twist for the process $eN\rightarrow e'h_1h_2X$ has 
been worked out in detail in Ref.~\cite{noi1}. Here, we will reconsider the case 
for an unpolarized beam and a transversely polarized target. The lab frame can be 
defined by the plane containing the beam 3-momentum $\vec l$ and the target 
polarization $\vec S$ (see Fig.~\ref{fig1}). The scattering plane, which contains 
the scattered lepton 3-momentum $\vec l'$ and $\vec l$, is rotated by the azimuthal 
angle $\phi_l^L$. Finally, the socalled hadronic plane, which contains $\vec q=\vec l
-\vec l'$ and $\vec P_h$, is rotated by $\phi_h^L = \phi_h^l+\phi_l^L$. A further 
plane, which is just sketched in Fig.~\ref{fig1} for sake of simplicity, contains 
$\vec P_1, \vec P_2, \vec R$ and, consequently, $\vec R_\perp$; it is rotated by 
$\phi_R^L=\phi_R^l + \phi_l^L$ with respect to the lab. 
The nine-fold differential cross section depends on the energy fraction taken by the
scattered lepton ($y=q^0/|\vec l|$), on $\phi_l^L$, on the quark light-cone 
momentum fraction $x=p^+/P^+$ of the target momentum $P$, on $z, \xi=z_1/z, 
\vec R_\perp$ and  $\vec P_{h\perp}$. Since $R_\perp^2 = \xi (\xi-1) P_h^2 -(1-\xi) 
P_1^2 -\xi P_2^2$~\cite{noi1}, the cross section can be more conveniently considered 
differential with respect to the pair invariant mass $P_h^2=M_h^2$ and $\phi_R^L$. 
By integrating over the ``internal'' dynamics (i.e. on $\xi, \vec k_\perp, 
\vec P_{h\perp}$) and by properly folding the cross section over the experimental 
set of beam and hadron-pair azimuthal positions $\phi_l^L$ and $\phi_R^L$, it is 
possible to come to the factorized expression
\begin{eqnarray}
\frac{ \langle d\sigma_{OT} \rangle }
     {dy \, dx \, dz \, dM_h^2}   
&\equiv &\int_0^{2\pi} d\phi_l^L \, d\phi_R^L \sin (\phi_R^L - 2 \phi_l^L ) 
    \int d\vec P_{h\perp} \, d\xi \, 
    \frac{   d\sigma_{OT}  }
         {dy \, d\phi_l^L \, dx \, dz \, d\xi \, dM_h^2 \, d\phi_R^L \, 
	   d\vec P_{h\perp}  }   \nonumber \\
&=  &\frac{ 4 \pi \alpha_{em}^2 s}{(2\pi)^3 Q^4} \, \sum_a e_a^2 x \, 
     \frac{(1-y) |\vec S_\perp |}{2(M_1+M_2)} \, 
     \int d\vec p_\perp \, h_1^a (x,\vec p_\perp^{\, 2}) \nonumber \\
&   &\quad \times \int d\xi \, |\vec R_\perp | \int_0^{2\pi} d\phi_R^L \int d\vec 
k_\perp \, H_1^{\open \, a} (z,\xi,M_h^2, \vec k_\perp^2, \vec k_\perp \cdot 
\vec R_\perp) \nonumber \\
&=  &\frac{\alpha_{em}^2 s}{4\pi^2 Q^4} \, \frac{(1-y) |\vec S_\perp |}{M_1+M_2} 
     \, \sum_a e_a^2 x \, h_1^a (x) \, H_1^{\open \, a}
     (z,M_h^2)
\label{eq:finalx}
\end{eqnarray}
where $\alpha_{em}$ is the electromagnetic fine structure constant, 
$s=Q^2/xy=-q^2/xy$ is the total energy in the center-of-mass frame, $M_1,M_2$ are 
the masses of the two observed hadrons, and a sum over the flavor $a$ of each quark 
and antiquark with charge $e_a$ is performed (see also Ref.~\cite{physrep}). In 
general, the integrated function $H_1^{\open}$ will depend on both $z$ and $M_h^2$, 
contrary to the assumption proposed in Ref.~\cite{jaffe}. It is worth noting also 
that Eq.~(\ref{eq:finalx}) suggests a simpler measurement with respect to the case 
where only one hadron is detected. In fact, here only the azimuthal position 
$\phi_R^L$ of the pair is required, while in order to measure the Collins effect 
both the azimuthal angle and the $|\vec P_{h\perp}|$ hadron momentum need to be 
known (see Ref.~\cite{boer} and references therein).

\begin{figure}[t]
\begin{center}
\epsfig{file=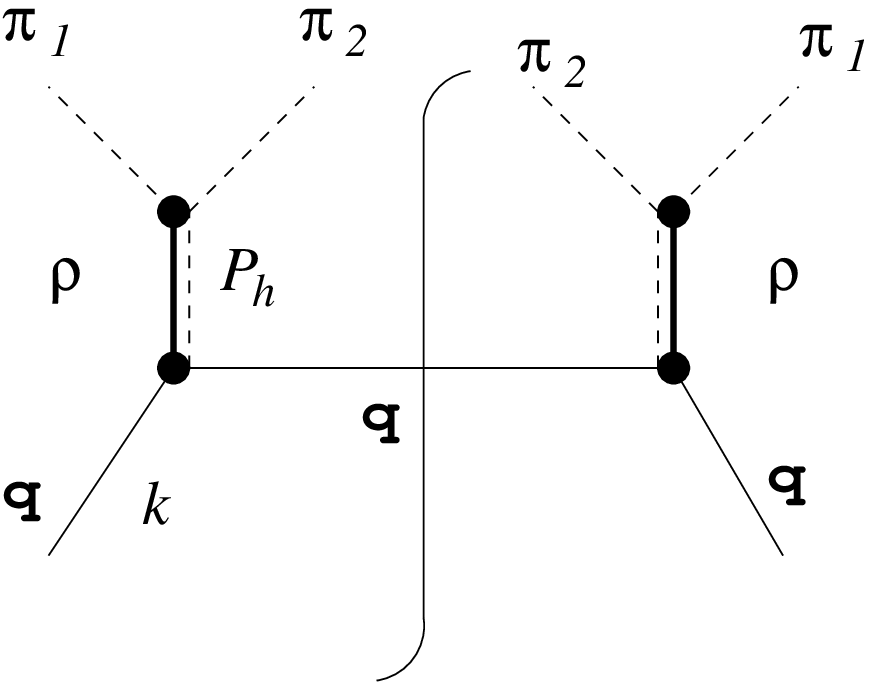,width=3.5cm,height=2.8cm}
\hspace{.5cm} 
\epsfig{file=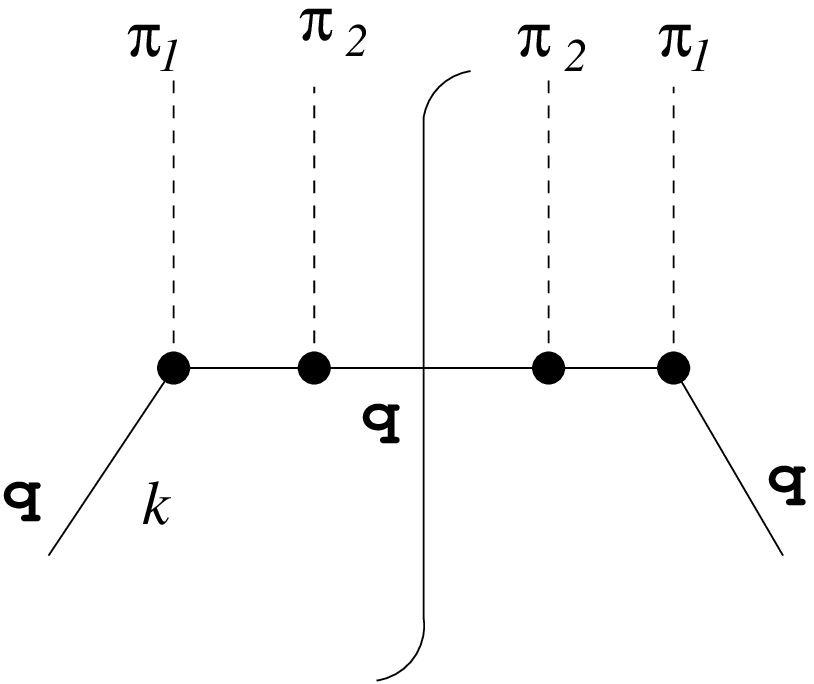,width=3.5cm,height=2.8cm} 
\hspace{.5cm}
\epsfig{file=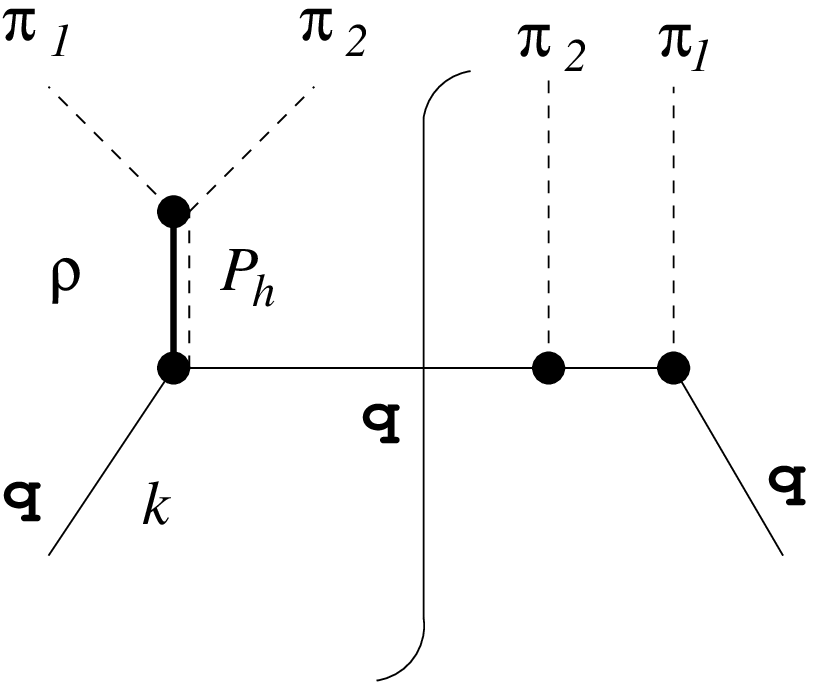,width=3.5cm,
height=2.8cm} 
\hspace{.3cm}
\begin{minipage}{3cm}
\epsfig{file=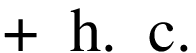,width=2cm}
\vspace{2.5cm}
\end{minipage}    \\[-1cm]
 a \hspace{4cm} b \hspace{4cm} c \hspace{2cm} {}
\end{center}
\vspace{-1cm}
\caption{The diagrams included for semi-inclusive detection of two pions with
invariant mass within the $\rho$ resonance width and in the spectator approximation. 
\label{fig2}}
\end{figure}

\begin{figure}[ht]
\begin{center}
\epsfig{file=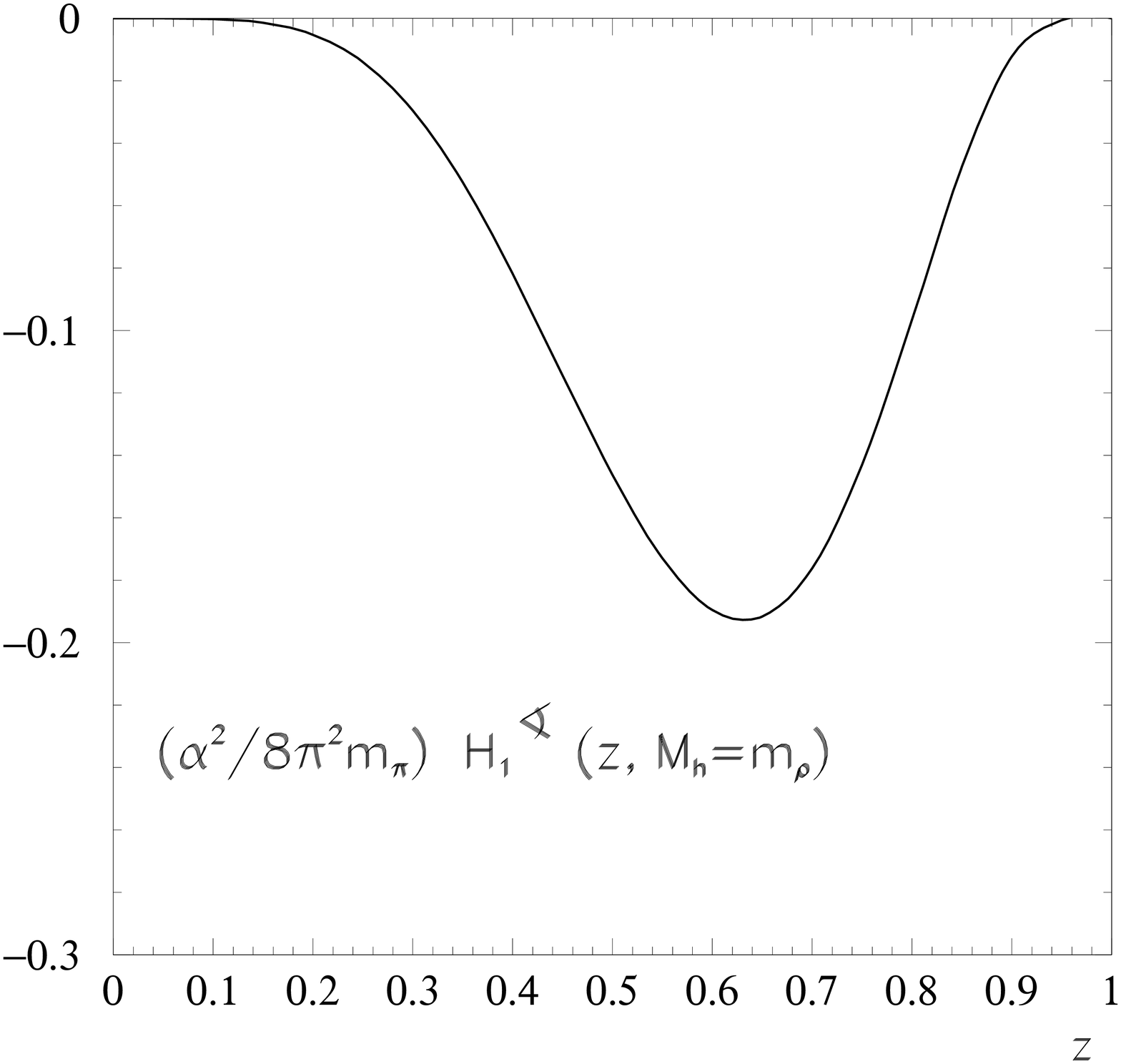,width=9cm,height=6.3cm}
\end{center} 
\vspace{-1.3cm}
\caption{The fragmentation function $\alpha_{em}^2 / [8\pi^2 m_\pi] \times 
H_1^{\open} (z,M_h^2=m_\rho^2)$ of Eq.~(\ref{eq:finalx}) for the 
$u\rightarrow \pi^+ \pi^-$ case. \label{fig3}}
\end{figure}

Quantitative predictions for $H_1^{\open \, a}$ in Eq.~(\ref{eq:finalx}) can be 
produced by extending the spectator model of Ref.~\cite{jakob} to the case of the 
emission of a hadron pair. For the hadron pair being a proton and a pion, results 
have been published in Ref.~\cite{noi2}, where FSI arise from the interference 
between the direct production and the Roper decay. Here, results are shown for the 
case of two pions with invariant mass in the range $[m_\rho-\Gamma_\rho, 
m_\rho+\Gamma_\rho]$, with $m_\rho=768$ MeV and $\Gamma_\rho\sim 250$ MeV. The 
spectator state has the quantum numbers of an onshell quark with constituent mass 
$m_q=340$ MeV. The quark decay at leading twist with the minimal number of vertices 
is represented by the set of diagrams shown in Fig.~\ref{fig2}, where now T-odd 
FF arise from the contribution of diagram \ref{fig2}c through the interference 
between the direct production of the two $\pi$ and the $\rho$ decay. Explicit check
has been made that the direct diagrams \ref{fig2}a and \ref{fig2}b qualitatively 
reproduce the experimental strength in the $P$- and $S$-channels, respectively, 
which represent most of the total $\pi - \pi$ production strength. 
In Fig.~\ref{fig3}, $\alpha_{em}^2 / [8\pi^2 m_\pi] \times H_1^{\open} 
(z,M_h^2=m_\rho^2)$ of Eq.~(\ref{eq:finalx}) is shown for the case $u\rightarrow 
\pi^+ \pi^-$. It shows that a nonvanishing integrated interference FF survives
allowing for the extraction of $h_1$ at leading twist, as suggested by 
Eq.~(\ref{eq:finalx}).

\vspace{-1.2cm}
\vspace{-0.5cm}

It is interesting to note that charge conservation for the diagrams of 
Fig.~\ref{fig2} implies that the same results are obtained for
the processes $u\rightarrow \pi^+ \pi^-$ and $d\rightarrow \pi^- \pi^+$, which
differ just by the interchange $\vec R_\perp \leftrightarrow - \vec R_\perp$.
Therefore, the $\vec k_\perp$-integrated amplitudes for the $u$ and $d$ quarks
leading to the same $\pi^+ \pi^-$ final state cancel each other. If only the valence
quark content is considered for the proton ($p$) and neutron ($n$) targets, then the
 cross section at leading twist for the $ep \rightarrow e'\pi^+ \pi^- X$ process 
turns out the same as for the $en \rightarrow e'\pi^- \pi^+ X$ one. Experimental
tests of this conjecture could shed light both on the validiy of the spectator
approximation and on the kinematical range where only the leading twist can be 
safely considered.


\section*{ACKNOWLEDGMENTS}
This work has been done in collaboration with A. Bianconi (Univ. Brescia) and R. 
Jakob (Univ. Wuppertal). Enlighting discussions with S. Boffi (Univ. Pavia) and 
D. Boer (RIKEN-BNL) are greatfully acknowledged.

\end{document}